\documentclass[twocolumn,trackchanges]{aastex701}

\usepackage{CJKutf8}

\usepackage{graphicx}
\usepackage[caption=false]{subfig}
\usepackage{tabularx}
\usepackage{capt-of}
\usepackage{hhline}
\usepackage{lineno}
\usepackage{amsmath}	
\usepackage{amssymb}	
\usepackage{textcomp}
\usepackage{upgreek}
\usepackage{listings}
\newcommand{\Msun}{\mathrm{M}_{\odot}}

\newcommand{\cmark}{\ding{51}} 
\newcommand{\xmark}{\ding{55}} 
\usepackage{pifont} 

\begin{document}

\title{Primordial Black Holes as Seeds for Extremely Overmassive AGN Observed by JWST}

\author[orcid=0000-0003-1541-177X,gname=Saiyang,sname=Zhang]{Saiyang Zhang\begin{CJK*}{UTF8}{bsmi}（張賽暘）\end{CJK*}}
\affiliation{Department of Physics, University of Texas at Austin, Austin, TX 78712, USA}
\affiliation{Weinberg Institute for Theoretical Physics, Texas Center for Cosmology and Astroparticle Physics, \\ University of Texas at Austin, Austin, TX 78712, USA}
\email[show]{szhangphys@utexas.edu}

\author[orcid=0000-0002-4966-7450,gname=Boyuan, sname=Liu]{Boyuan Liu\begin{CJK*}{UTF8}{bsmi}(劉博遠)\end{CJK*}} 
\affiliation{Institute of Astronomy, University of Cambridge, Madingley Road, Cambridge, CB3 0HA, UK}
\affiliation{Universit\"at Heidelberg, Zentrum fur Astronomie, Institut f\"ur Theoretische Astrophysik, D-69120 Heidelberg, Germany}
\email[show]{boyuan.liu@uni-heidelberg.de}

\author[orcid=0000-0003-0212-2979,gname=Volker,sname=Bromm]{Volker Bromm}
\affiliation{Weinberg Institute for Theoretical Physics, Texas Center for Cosmology and Astroparticle Physics, \\ University of Texas at Austin, Austin, TX 78712, USA}
\affiliation{Department of Astronomy, University of Texas at Austin, Austin, TX 78712, USA}
\email{vbromm@astro.as.utexas.edu}

\author[orcid=0000-0002-1528-1920,gname=Florian,sname=K\"uhnel]{Florian K\"uhnel}
\affiliation{Max-Planck-Institut f{\"u}r Physik, Boltzmannstr.~8, 85748 Garching, Germany}
\affiliation{Arnold Sommerfeld Center, Ludwig-Maximilians-Universit{\"a}t, Theresienstr.~37, 80333 M{\"u}nchen, Germany}
\email{Florian.Kuehnel@physik.uni-muenchen.de}


\begin{abstract}
The James Webb Space Telescope (JWST) has recently identified Abell 2744–QSO1 as a compact, metal-poor, black hole (BH) dominated galaxy at $z\simeq 7$. This system exhibits an extreme black-hole-to-stellar mass ratio and unusually low metallicity, posing significant challenges to BH seeding models. Motivated by these discoveries, we perform high-resolution cosmological simulations with a massive primordial black hole (PBH; $M_{\rm BH}=5\times10^7\,M_\odot$) seed, incorporating for the first time a fully coupled treatment of PBH accretion, BH feedback, and Population~III/II star formation and stellar feedback. Although PBHs accelerate structure formation through the seed effect, the associated strong thermal feedback from the accretion delays the onset of star formation to $z\lesssim 10$, producing short, bursty episodes throughout the subsequent evolution. 
PBH-driven outflows expel enriched gas from the nucleus, while sustained inflows from the intergalactic medium continuously replenish pristine material. This feedback-regulated cycle naturally yields low accretion rates ($\dot{m}_{\rm BH}/\dot{m}_{\rm edd} \sim 1-10\%$), subsolar metallicities ($Z/Z_\odot\lesssim10^{-2}$) and extreme $M_{\rm BH}/M_\star$ ratios during both the initial star-forming phase and the subsequent quenching phases, in excellent agreement with JWST observations. Our results demonstrate that massive PBHs offer a viable pathway for forming the most extreme high-redshift systems, providing a physically motivated explanation for the extraordinary properties of Abell 2744–QSO1, as a sub-class of the broader population of JWST-discovered ``little red dots’’.

\end{abstract}


\keywords{\uat{Dark matter}{353} --- \uat{Early universe}{435} --- 	\uat{Galaxy formation}{595} --- \uat{Population III stars}{1285} --- \uat{Supermassive black holes}{1663}}


\section{Introduction}

Since the launch of the \textit{James Webb Space Telescope} (JWST), an unexpected population of massive supermassive black holes (SMBHs) has been revealed at very high redshifts ($z \gtrsim 7$)~\citep[e.g.,][]{Goulding2023ApJ...955L..24G, Larson_2023_BH, Bogdan:2023UHZ1,Greene2024,Kovacs2024ApJ...965L..21K, Maiolino2024A&A,Natarajan:2023UHZ1,GHZ9Napolitano2025ApJ...989...75N}. Among these early systems, a class of compact, extremely red, and spectroscopically peculiar sources, commonly referred to as ``little red dots'' (LRDs), has emerged as a distinct population \citep[e.g.,][]{Labbe2023Natur.616..266L, KokorvLRD2024ApJ...968...38K,Leung2024:LRD, KocevskiLRD2025ApJ...986..126K,Taylor2025ApJ...986..165T,LabbeLRD2025ApJ...978...92L}. These newly-discovered objects are found to host central black holes (BHs) that are often `overmassive' relative to their stellar components, with black-hole-to-stellar mass ratios exceeding the local relation~\citep[e.g.][]{PacucciLRD2023ApJ...957L...3P,Inayoshi2024ApJ...966..164I}. Among those discoveries, one of the most striking examples is the JWST source Abell~2744--QSO1, whose inferred BH mass ($M_{\rm BH} \sim 5 \times 10^7\,M_\odot$), lack of significant stellar contributions in its kinematics and spectra ($M_\star\lesssim 2\times 10^7\ \rm M_\odot$), combined with extremely low metallicity ($Z \lesssim 0.01\,Z_\odot$) and low accretion efficiency ($f_{\rm Edd} \sim 0.01 - 0.03$) at $z \sim 7$ pose severe challenges to conventional models of SMBH formation and coevolution with its host galaxy~\citep{FurtakAbell27442023ApJ,BlackTHUNDER2025,Maiolino2025arXiv250522567M, QSO1Direct2025arXiv250821748J}.

Traditional light-seed scenarios, typically related to stellar remnants of Population~III (Pop~III) stars~\citep[see reviews in, e.g.,][]{Smith2019:BHreview, Inayoshi:2020}, struggle to reproduce both the high inferred accretion rates and the black-hole-to-stellar mass ratios implied by JWST observations~\citep[e.g.,][]{Milosavljevic2009, Jeon2023, Bogdan:2023UHZ1}. These models generally require sustained super-Eddington accretion over extended periods~\citep[e.g.,][]{Alexander2014Sci...345.1330A, Jeon2025ApJ...979..127J}.
An alternative pathway is the formation of massive BHs through the direct collapse of pristine, metal-poor gas with minimal fragmentation~\citep{Loeb1994ApJ...432...52L, BrommDCBH2003ApJ...596...34B, Begelman2006:DCBH, Lodato:2006DCBH}. Such conditions typically demand a nearby source of intense soft-UV, Lyman–Werner (LW) radiation to suppress molecular hydrogen cooling~\citep[e.g.,][]{AgarwalDCBH2014MNRAS.443..648A, Habouzit2016, OBrennan2025}, which is absent in the vicinity of Abell~2744--QSO1~\citep{QSO1Direct2025arXiv250821748J}.

 Another explanation involves the presence of \textit{primordial black holes} (PBHs) that predate galaxy formation~\citep[for analytical work see, e.g.,][]{DuchtingPBH2004PhRvD..70f4015D, Bernal:2017nec, DeLuca2023PhRvL.130q1401D, Dayal2024A&A...690A.182D, DeLuca2025arXiv251219666D}~\footnote{A few alternative formation scenarios have been proposed, such as core collapse of strongly self-interacting dark matter~\citep[e.g.,][]{Jiang2025SIDM} or direct collapse of gas at $z\gtrsim 100$ from the enhanced power spectrum at small scales~\citep[e.g.,][]{Qin2025arXiv250613858Q}}. PBHs, hypothesized to form from the collapse of density fluctuations shortly after the Big Bang~\citep{Zeldovich1967SvA....10..602Z, hawking1971gravitationally, Carr1975ApJ...201....1C, Belotsky2019, Escriva2022}, may constitute only a small fraction of the dark matter, yet can profoundly influence cosmic evolution~\citep[for a review, see][]{Carr2020ARNPS..70..355C}.
On larger scales, depending on their mass spectrum and abundance, PBHs can enhance primordial density fluctuations and seed the formation of dark matter halos~\citep[e.g.,][]{Mack2007ApJ, Ricotti2007ApJI, Ricotti2008ApJII, Carr2018MNRAS.478.3756C, Inman2019PhRvD.100h3528I, Su2023, Colazo2024, Zhang:2024PBH}, thus accelerating early structure formation~\citep{Meszaros1975A&A....38....5M, Afshordi2003ApJ...594L..71A, Kashlinsky2021PhRvL.126a1101K, Cappelluti2022ApJ, Boyuan2022ApJ}. 
Their early accretion can heat and ionize the intergalactic medium (IGM), thus modifying cosmic radiation backgrounds~\citep[e.g.,][]{Ali-Haimoud2017PhRvD, Deluca2020JCAP...06..044D, Lu2021, Takhistov2022, Ziparo2022, Zhang2024MNRAS.528..180Z, Casanueva-Villarreal2024}, and potentially trigger the first episodes of star formation around PBH-seeded halos~\citep{Boyuan2022MNRAS.514.2376L, Boyuan2023arXiv231204085L}.
Recent studies further suggest that massive PBHs ($\gtrsim10^6\,M_\odot$) can induce the formation of bound stellar clusters~\citep{Zhang2025PBHStar} or secondary massive BHs~\citep{Zhang2025pbhdcbh}, giving rise to systems reminiscent of the LRDs discovered by JWST~\citep{DayalPBH2025arXiv250608116D,Matteri2025PBH}.
However, most previous simulations have included only BH accretion and feedback while ignoring stellar feedback (or vice versa), leaving open how PBHs coevolve with PBH-seeded galaxies in the early Universe.

In this work, we integrate the legacy stellar feedback model~\citep{Jaacks2018, Jaacks2019, Liu2020MNRAS.497.2839L,Liu2020} into our cosmological PBH simulation framework. This approach enables, for the first time, a combined treatment of both BH and stellar feedback in the early growth of PBH-seeded systems. 
Our simulations follow the coupled evolution of dark matter, baryons and stellar components from $z \simeq 1100$ to $z \simeq 7$, reproducing the physical conditions of Abell~2744--QSO1 and potentially other JWST LRDs, aiming to assess whether heavy PBH seeds can naturally account for the formation of these extreme high-redshift systems.

In Section~\ref{sec:method}, we describe the numerical setup for the simulation, focusing on the recipes for star formation and stellar feedback. From the simulated results, the detailed galaxy assembly and metal enrichment history around the central PBH is discussed in Section~\ref{sec:Results}. The limitations and caveats are addressed in Section~\ref{sec:Caveats}, and we present our conclusions in Section~\ref{sec:conclusion}.

\begin{table*}[htb!]
    \centering
    \caption{Summary of key parameters and main results. 
    $z_{\rm ini}$ the initial redshift where the simulation starts, and $N_{\rm eff}$ the total number of particles within the simulation box of size $L = 1\,h^{-1}\,\mathrm{Mpc}$. $\epsilon_r$ is the thermal coupling efficiency of BH feedback, $\langle f_{\rm edd}\rangle \equiv \langle\dot{m}_{\rm BH}/\dot{m}_{\rm edd}\rangle $ is the time-averaged BH accretion rate in units of the Eddington limit, \texttt{FULL\_SF} a flag to control whether we consider the full stellar feedback recipe or not, whereas  $\epsilon_{\rm DM}$, $\epsilon_{\rm gas}$, and $\epsilon_{\star}$ represent the (co-moving) softening lengths for PDM, gas, and stellar particles, respectively. Finally, $z_{\rm col}$ is the redshift when star formation begins around the central PBH, and $M_{\star}$ denotes the total mass of stars formed in the PBH-hosting halo, evaluated by the end of the simulation ($z \sim 7$). }
    {
    \begin{tabular}{cccccccccccc}
    \hline
        Run &$z_{\rm ini}$  & $N_{\rm eff}$  & $\epsilon_{\rm r}$ &$\langle f_{\rm edd}\rangle$  & \texttt{FULL\_SF}& $\epsilon_{\rm DM}\small[\rm kpc/h]$& $\epsilon_{\rm gas}\small[\rm kpc/h]$ & $\epsilon_{\star}\small[\rm kpc/h]$& $z_{\rm col}$&  $M_{\star}\small [\Msun]$ \\

    \hline
    
    \texttt{PBH\_DMonly}   & 3400  & $ 256^3$  & -& -& - & 0.05  & - & - & - & -\\
    \texttt{PBH\_M5e7\_fd005}   & 1100 &  $2\times 256^3$  &  0.005 & 0.0076& \xmark & 0.05 & 0.05 & 0.005 & 9.64& $2.0\times 10^7$   \\
    \texttt{PBH\_SF\_M5e7\_fd005}   & 1100 &  $2\times 256^3$  &  0.005 & 0.0081 & \cmark & 0.05 & 0.05 & 0.005  & 8.20 &  $7.7\times 10^5$\\
  \texttt{PBH\_SF\_M5e7\_fd0025}   & 1100 &  $2\times 256^3$  & 
     0.0025 & 0.013 & \cmark & 0.05 & 0.05 & 0.005  & 8.58 & $3.2\times 10^5$\\
    
    \texttt{PBH\_SF\_M5e7\_fd01}   & 1100 &  $2\times 256^3$  &  
    0.01& 0.0047  & \cmark & 0.05 & 0.05 & 0.005  & 12.80 &  $1.9\times 10^6$\\
    \hline
    \end{tabular}
    }
    \label{Table:SimParam}
\end{table*}

\section{Methodology} \label{sec:method}

The simulation is performed using the \textsc{GIZMO} code~\citep{Hopkins2015MNRAS.450...53H}, which couples the Lagrangian meshless finite-mass (MFM) hydrodynamics solver (with $N_{\rm ngb} = 32$ neighbors) to the parallel Tree+PM gravity solver of \textsc{GADGET-3}~\citep{springel2005cosmological}. The non-equilibrium primordial chemistry and cooling network includes 12 species, following \citet{Bromm2002ApJ...564...23B} and \citet{Johnson2006MNRAS.366..247J}, supplemented by C~\textsc{II}, O~\textsc{I}, Si~\textsc{II} and Fe~\textsc{II} line cooling~\citep{Jaacks2018}, with initial abundances at $z \simeq 1100$ set by \citet{Galli2013ARA&A..51..163G}. The \textit{Planck18} cosmological parameters were adopted throughout \citep{Plank2020A&A...641A...6P}: $\Omega_{\rm m} = 0.3111$, $\Omega_{\rm b}=0.04897$, $h = 0.6776$, $\sigma_8 = 0.8102$, $n_{\rm s} = 0.9665$.

We describe the initial conditions and general numerical setup in Section~\ref{subsec:ini}, and the stellar formation and feedback prescriptions in Section~\ref{subsec:SF}. The relevant parameters for all runs are summarized in Table~\ref{Table:SimParam}.

\subsection{Initial Conditions}\label{subsec:ini}

To reproduce the observed properties of Abell~2744--QSO1, we focus on the cosmic evolution surrounding an isolated PBH. The simulation box has a comoving side length of $L = 1\,h^{-1}\,\mathrm{Mpc}$ and contains $N_{\rm PDM} = 256^3$ dark matter particles. The simulation begins at $z \simeq 3400$, where we place a BH of mass $5\times10^7\,M_\odot$ at the center of the box~\footnote{In this work, we simulate structure formation around an isolated PBH, motivated by the inferred mass of Abell~2744--QSO1~\citep{QSO1Direct2025arXiv250821748J}.  The chosen box size limits the effective PBH mass fraction to $f_{\rm PBH} \lesssim 8\times10^{-4}$, which is broadly compatible with formation scenarios based on early-Universe phase transitions~\citep{Carr2021PDU....3100755C}. However, such a massive PBH could also originate from mergers of initially clustered lower-mass PBHs~\citep{Zhang:2024PBH, ZhangPBHCluster2025arXiv250707171Z}.}, generated using the \textsc{MUSIC} code~\citep{hahn2011multi}.
The numerical procedure for embedding the PBH and perturbing the surrounding particle dark matter (PDM) field is implemented using the \textsc{PHANTOM} code~\citep{zhang_2025_17025634}, following the formalism in \citet{Ali-Haimoud2017PhRvD, Inman2019PhRvD.100h3528I, Boyuan2022MNRAS.514.2376L, Zhang:2024PBH}, providing the initial condition for the \texttt{PBH\_DMonly} run. We first evolve this run from $z \simeq 3400$ to the recombination epoch ($z \simeq 1100$) to establish the PDM distribution around the central PBH~\footnote{The PBH is treated as a freely moving collisionless particle providing the central gravitational potential; numerically, it is evolved like any other particle except for a boosted drag term during accretion to minimize BH wandering. To be specific, we apply a drag force $\vec{F}=-\dot{m}_{\rm edd}\vec{v}_{\rm rel}$ to the BH, given its Eddington accretion rate $\dot{m}_{\rm edd}$ and velocity $\vec{v}_{\rm rel}$ relative to gas. Consequently, the BH-gas relative velocity remains well below the local sound speed throughout the simulation, rendering its contribution to the accretion rate subdominant.}. We then add a uniform baryonic component to initialize the full hydrodynamical runs, resulting in a total of $N = 2\times256^3$ particles for both the \texttt{PBH\_SF\_M5e7\_fd005} and \texttt{PBH\_M5e7\_fd005} runs with and without stellar feedback, respectively. The resulting gas and dark matter particle masses are $1.2\times10^3\,M_\odot$ and $5.1\times10^3\,M_\odot$. 

BH accretion and feedback are modeled using the same subgrid prescription as in \citet{Zhang2025PBHStar} which self-consistently tracks the growth of the central PBH through Bondi-Hoyle-Lyttleton accretion with radiative/thermal feedback. The feedback is implemented by depositing energy into the surrounding gas according to a fiducial thermal radiation coupling efficiency of $\epsilon_r = \delta E_{\bullet, \rm inject} / (L_{\bullet, \rm acc} \delta t) = 0.005$~\footnote{The adopted feedback coefficient is also chosen to reproduce the observed properties of the QSO-1 system, allowing for moderate star formation consistent with its inferred stellar content. We have also included two additional runs ($\epsilon_r = 0.0025$, \texttt{PBH\_SF\_M5e7\_fd0025} and  $\epsilon_r =0.01$, \texttt{PBH\_SF\_M5e7\_fd01}) by varying this parameter (see Table~\ref{Table:SimParam}). In this work, we do not explicitly model UV/LW radiation from BH accretion, given the limited understanding of the BH intrinsic SED in the early universe~\citep[see, e.g.,][]{Greene2026ApJ...996..129G}. We refer the reader to our previous work \citep{Zhang2025PBHStar, Zhang2025pbhdcbh}, and to a forthcoming study for a more comprehensive discussion of accretion feedback and the assumed seed mass.}, where $\delta E_{\bullet, \rm inject}$ represents the thermal energy injected into the ambient medium, and $L_{\bullet, \rm acc}$ is the accretion luminosity. Both runs are evolved from $z \simeq 1100$ to $z \simeq 7$, corresponding to the observed redshift of Abell~2744--QSO1~\citep{BlackTHUNDER2025}.

    \vspace{-2ex}

\subsection{Star Formation and Stellar Feedback}\label{subsec:SF}

During the simulation, when a gas particle becomes Jeans unstable near the central BH and survives for a free fall time, without being accreted by the BH or dispersed by feedback, it is converted into a stellar sink particle according to the star formation criterion of \citet{Zhang2025PBHStar}. Our analysis focuses primarily on the \texttt{PBH\_SF\_M5e7\_fd005} run, which includes a physically motivated stellar evolution and feedback model, unlike the \texttt{PBH\_M5e7\_fd005} run in which sink particles remain passive. 

A metallicity threshold of $Z_{\rm th} = 10^{-4}\,Z_\odot$ (with $Z_\odot = 0.02$) distinguishes Pop~III and Population~II (Pop~II) star formation. Pop~III stars follow a modified Larson IMF, where $dN / dM \propto M^{-\alpha} \exp(-M_{\rm cut}^2 / M^2)$, with $\alpha = 0.17$, $M_{\rm cut}^2 = 20 ~\Msun^2 $, over the mass range $1-150 \, \Msun$~\citep{Jaacks2018}, whereas a Chabrier IMF with a mass range of $0.08 - 100\,\Msun$ is assumed for Pop~II stars~\citep{Jaacks2019}.  
Each stellar particle (Pop~III or Pop~II) has a mass of $\simeq 600\,M_\odot$. For Pop~III, this mass corresponds to the typical mass of Pop~III star clusters formed along the standard $\rm H_2$-cooling pathway~\citep[e.g.,][]{Stacy2013PopIII,Hirano2017MNRAS.470..898H,Liu2021,Liu2024_Mass}. 

Stellar feedback includes ionization heating, photo-dissociation of $\rm H_2$ by Lyman-Werner (LW) radiation, thermal energy injection and metal enrichment from supernovae~\citep{Jaacks2018, Jaacks2019}. 
The LW radiation field is computed based on both the global star formation rate density and local contributions from stellar particles~\citep[]{Liu2020}, assuming an optically thin regime while accounting for self-shielding~\footnote{Here we do not consider LW radiation from the central BH (explored in~\citealt{Zhang2025PBHStar,Zhang2025pbhdcbh}).}. 
Reionization heating of the intergalactic medium (IGM) is modeled through a uniform global UV background with a characteristic self-shielding scale of $\sim1$~kpc, based on the photoionization rate calculated by \citet{FaucherUVB2009}.  
The supernova (SN) feedback model follows the implementation of \citet{Jaacks2019}, in which each stellar particle injects thermal energy, metals, and momentum into the surrounding gas based on IMF-averaged yields. For a typical stellar particle of mass $m_\star \simeq 600~\Msun$, Pop~III explosions release of order $\sim 7\times10^{51}\,\mathrm{erg}$ with metal yields $\sim 39~\Msun$, while Pop~II particles contribute $\sim 6.7\times10^{51}\,\mathrm{erg}$ and $\sim 10~\Msun$ of metals, with lifetimes of $\sim 3$ and $\sim 10$~Myr, respectively. The resulting SN bubbles expand to characteristic radii of $\sim 650$~pc. A detailed description of the adopted parameters and their derivation can be found in Table~1 of \citet{Liu2020MNRAS.497.2839L} and in the feedback model presented by \citet{Jaacks2018, Jaacks2019}. 


    

\section{Results} \label{sec:Results}
With the stellar feedback recipes described in Section~\ref{sec:method}, we now analyze the resulting PBH simulation runs, comparing to previous work\footnote{Also see the appendix in~\cite{Maiolino2025arXiv250522567M}.} that did not include the full feedback physics. To study the synergy between stellar and BH feedback, metal enrichment, and gas outflow, we first track the star formation history (SFH) and the assembly of galaxies in Section~\ref{subsec:SFH}, followed by a discussion of the metal enrichment history in Section~\ref{subsec:Metal}.


\begin{figure}[htb!]
    \centering
\vspace{-5ex}
\includegraphics[width=0.95\columnwidth]{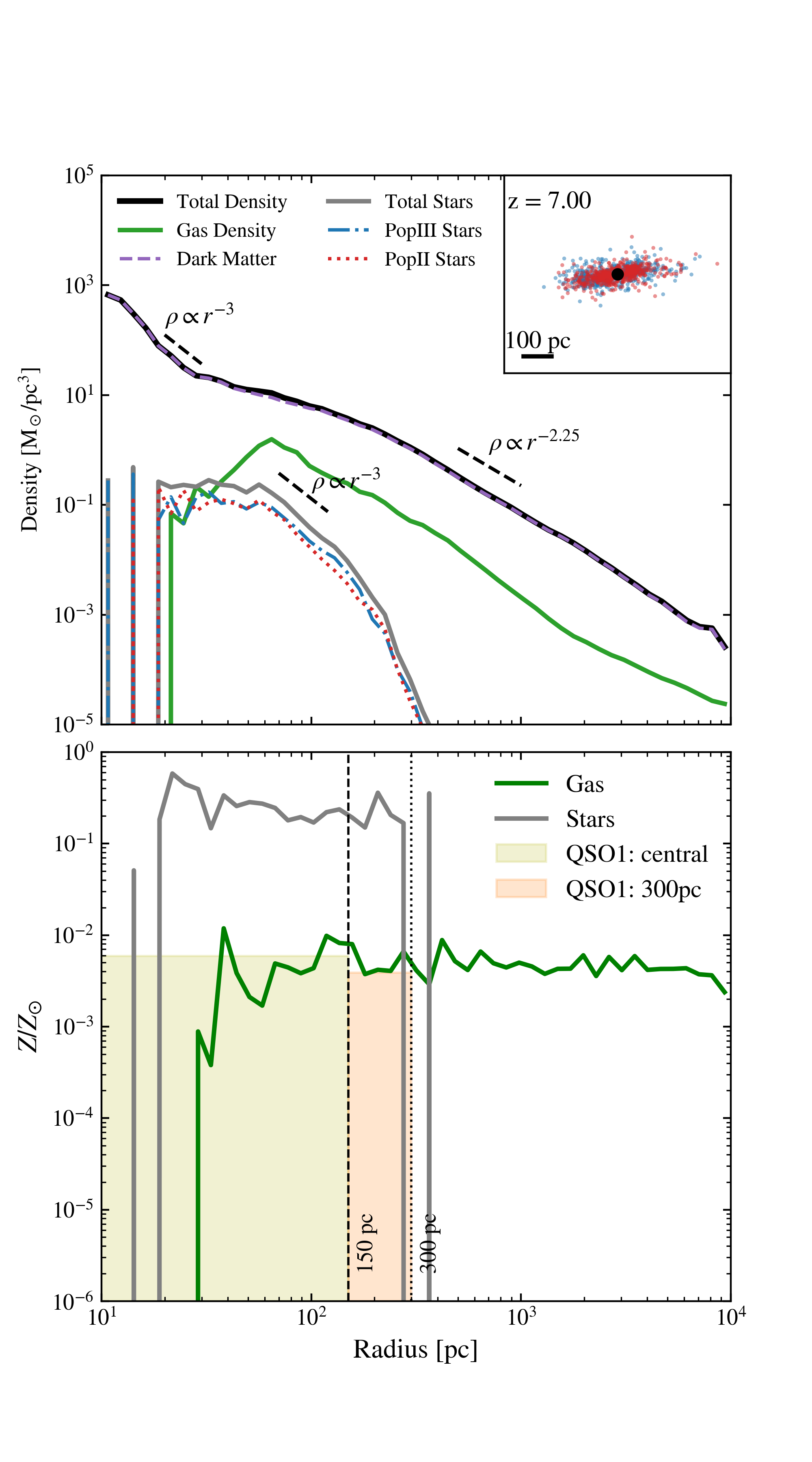}
\vspace{-8ex}
    \caption{Matter density and metallicity profiles around the central BH at $z=7$. Shown are the spherically averaged density and metallicity profiles from the final snapshot of the \texttt{PBH\_SF\_M5e7\_fd005} run. \textbf{Top:} Density of total matter (black), gas (green), dark matter (purple dashed), total stellar content (gray), Pop~III stars (blue dashed–dotted), and Pop~II stars (red dotted) vs. radius. Dashed black lines illustrate select power-law forms that match the simulated density profiles at certain parts. 
    The inset shows the projected spatial distribution of Pop~III (blue) and Pop~II (red) stars relative to the central PBH (black circle), with a scale as indicated. \textbf{Bottom:} Radial metallicity profile for gas (green) and stars (grey), expressed in units of solar metallicity $Z_\odot$. Shaded regions indicate the observed metallicity constraints for Abell~2744--QSO1 within $\sim150$~pc (yellow) and $\sim300$~pc (orange) apertures~\citep{Maiolino2025arXiv250522567M}. The simulated system exhibits a steep inner density profile and maintains subsolar gas metallicity throughout the central $\sim300$~pc, in agreement with the properties inferred for QSO1.}
    \label{fig:profileplot}
\end{figure}

\subsection{Star Formation History}\label{subsec:SFH}

In~\citet{Zhang2025PBHStar}, a BH seed of $10^6\,\Msun$ was placed at the center of the simulation box (with $L = 0.25\,h^{-1}\,\mathrm{Mpc}$), accelerating structure formation and seeding a $\sim10^8 \,\Msun$ halo by the termination redshift $z\sim 9$. When a fiducial feedback efficiency of $\epsilon_r = 0.005$ was assumed, star formation commenced at $z\sim30$, and the BH growth remained modest ($\Delta m_{\rm BH} / m_{\rm BH} \lesssim 5\%$) by $z\sim9$. In contrast, the present work embeds a much heavier initial seed of $5\times10^7 \,\Msun$, resulting in a halo that is $\sim 50$ times more massive at the same redshift\footnote{The halo mass seeded by a PBH is expected to scale linearly with the PBH mass~\citep{Mack2007ApJ}.}and providing a substantially larger gas reservoir available for accretion and star formation. At the same time, the rate of energy injection from BH accretion feedback scales as $ \delta E_{\bullet, \rm inject}/ \delta t = \epsilon_{\rm EM}\dot{m}_{\rm BH} c^2 \propto m_{\rm BH}^2$, implying that the more massive seed both consumes gas more rapidly and injects significantly greater energy into the surrounding medium than the $10^6\,\Msun$ case. 

\begin{figure}[htb!]
    \centering
    \includegraphics[width=\columnwidth]{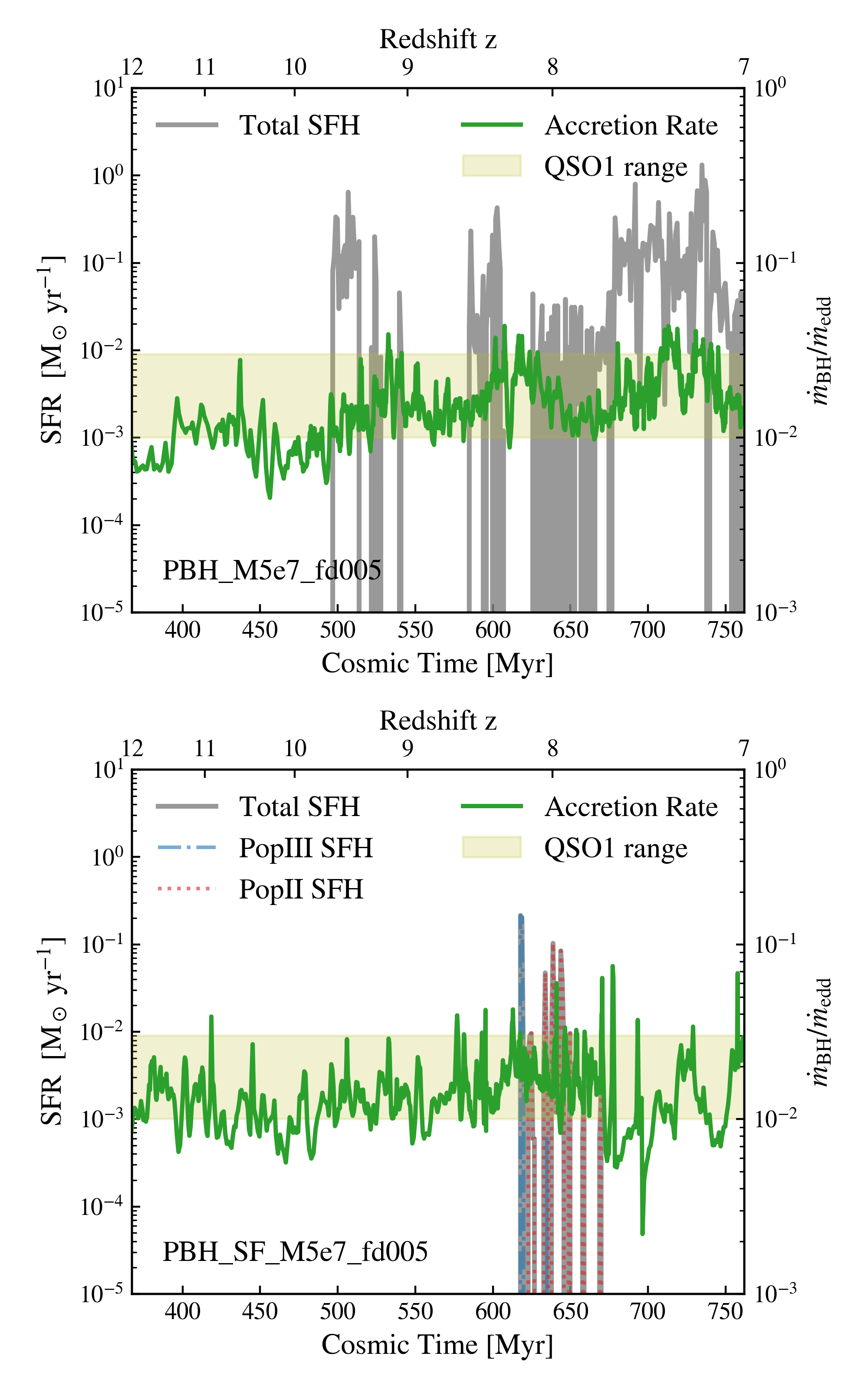}
    \vspace{-5ex}
    \caption{\textbf{Combined BH accretion and star formation history.} Shown are the BH accretion rates and star formation rates for the \texttt{PBH\_M5e7\_fd005} and \texttt{PBH\_SF\_M5e7\_fd005} runs. \textbf{Top:} Accretion rate (green), expressed as a fraction of the Eddington limit, $\dot{m}_{\rm BH}/\dot{m}_{\rm edd}$, together with the total star formation rate (gray) within the vicinity of the PBH for the \texttt{PBH\_M5e7\_fd005} run. Both quantities are shown as a function of cosmic time, with the corresponding redshift indicated on the upper axis. The shaded region (yellow) represents the accretion rate of Abell 2744-QSO1 inferred from observations as $\dot{m}_{\rm BH} / \dot{m}_{\rm edd}~\sim 1-3\%$~\citep{QSO1Direct2025arXiv250821748J}. \textbf{Bottom:} Same as the top panel, but for the \texttt{PBH\_SF\_M5e7\_fd005} run. In this case, the total SFR (gray) is decomposed into its Pop~III (blue dash-dotted) and Pop~II (red dotted) components. In both simulations, BH accretion remains at $\sim 1-10\%$ of the Eddington level, while star formation occurs in short, bursty episodes regulated by BH feedback. }
    \label{fig:SFH}
\end{figure}

In this study, for both \texttt{PBH\_M5e7\_fd005} and \texttt{PBH\_SF\_M5e7\_fd005}, the BH accretes at $\sim 1-10$\% of the Eddington rate, in good agreement with the Eddington ratio $\sim 0.01-0.03$ of Abell 2744-QSO1 inferred by JWST. In our simulations, the BH spends roughly half of its time accreting within this range (see Fig.~\ref{fig:SFH}), ultimately reaching a final mass of $\sim6\times10^7\,\Msun$ by $z=7$. This strong feedback delays the onset of star formation until $z\lesssim 10$, despite the ample gas supply near the BH (see Table~\ref{Table:SimParam}). We also present two additional cases, \texttt{PBH\_SF\_M5e7\_fd0025} and \texttt{PBH\_SF\_M5e7\_fd01}, where $\epsilon_r$ is reduced and enhanced by a factor of 2, respectively. We found that varying the feedback efficiency does not translate into a systematic shift of the star formation redshift, as both reducing and enhancing $\epsilon_r$ with respect to the fiducial choice leads to an earlier onset of star formation. Similarly, the central gas metallicity does not show any clear trend with $\epsilon_r$, nor do we find a direct scaling with the total stellar mass formed. This reflects the stochastic/bursty nature of early star formation and metal mixing in PBH-seeded halos. 
Lowering $\epsilon_r$ leads to a higher time-averaged BH accretion rate, approximately following\footnote{This empirical scaling is obtained from an extended suite of runs spanning multiple values of $\epsilon_r$. For brevity, we do not show all runs in this paper, but the quoted fit is based on the full set.} $\langle\dot{m}_{\rm BH}/\dot{m}_{\rm edd}\rangle\propto \epsilon_r^{-0.9}$, which partially compensates for the reduced feedback efficiency in maintaining a comparable time-averaged energy injection rate ($\propto\epsilon_r \dot{m}_{\rm BH}$) across different simulations. As a result, the net heating of the gas 
only weakly depends on $\epsilon_r$, and the relevant effect is overwhelmed by stochasticity, such that changing $\epsilon_r$ does not necessarily decrease nor delay the onset of gas collapse in a particular halo. 

We terminate the simulation at $z\sim7$ and analyze the resulting stellar and gaseous structures. Figure~\ref{fig:profileplot} shows the radial distribution of stars, gas, dark matter, and metals surrounding the central PBH in the final snapshot of the \texttt{PBH\_SF\_M5e7\_fd005} run, while Figure~\ref{fig:SFH} presents the star formation histories for both the \texttt{PBH\_M5e7\_fd005} and \texttt{PBH\_SF\_M5e7\_fd005} simulations. The density profile in Figure~\ref{fig:profileplot} exhibits a compact, approximately flat stellar core ($\rho\propto r^0$) within $r\lesssim r_{1/2}\sim 50\ \rm pc$, transitioning to a steep decline ($\rho\propto r^{-3}$) at $r\gtrsim50$~pc where the gas component dominates. The overall dynamics, however, is governed by the dark matter halo, which develops a pronounced spike at $r\lesssim30$~pc ($\rho\propto r^{-3}$) and a shallower profile at larger radii ($\rho\propto r^{-2.25}$ for $r\gtrsim100$~pc). A similar structure is found in the \texttt{PBH\_M5e7\_fd005} run, though the absence of stellar feedback results in a more centrally concentrated stellar component. The total enclosed mass of dark matter, gas, and stars (as extended components) within $150\ \rm pc$ is comparable to the BH mass, generally consistent with the resolved kinematics\footnote{We note that the mass ratio of the point source and the extended component inferred from observations depends on the density profile assumed for the latter. Besides, the dark-matter spike within $\lesssim30$~pc cannot be separated from the BH point source observationally. Combining the spike with the BH as an effective central point mass, we obtain the simulated ratio $M_{\rm point}/M_{\rm extended}\approx2.2$ (0.93) within 100 (150)~pc, in agreement with the lack of significant contribution of an extended component in the resolved kinematics of QSO1, which indicates that $M_{\rm point}/M_{\rm extended}\gtrsim 2$, assuming a $n = 1$ Sérsic density profile of half mass radius 100~pc truncated at 300~pc for the extended component \citep{QSO1Direct2025arXiv250821748J}.} of Abell~2744--QSO1 in observations~\citep{QSO1Direct2025arXiv250821748J}.

Both simulations exhibit strongly episodic star formation. The initial burst is dominated by Pop~III stars, reaching a peak rate of $\sim1\ \Msun\,\mathrm{yr}^{-1}$ before transitioning into Pop~II star formation. In the \texttt{PBH\_M5e7\_fd005} run, this first burst lasts for $\sim50$~Myr, 
 which is followed by a $\sim50$~Myr quiescent period before renewed star formation occurs. By $z\sim 7$, a stellar mass of $\sim 2\times 10^7 \,\Msun$ has formed near the PBH, close to the upper limit for Abell~2744--QSO1 placed by kinematic analysis. In contrast, the inclusion of stellar feedback in the \texttt{PBH\_SF\_M5e7\_fd005} run drastically reduces the star formation efficiency: only a single $\sim50$~Myr episode occurs, after which the system is quenched. By $z\sim7$, a total stellar mass of $\sim 7.7\times10^5\,\Msun$ ($4.2\times 10^5\,\Msun$\,
in Pop~III and $3.5\times 10^5\,\Msun$ 
in Pop~II stars) has formed in the vicinity of the BH, assembling into a star cluster with a half-mass radius of $r_{1/2}\sim 55$~pc
and a peak density exceeding $0.1\,\Msun\,\mathrm{pc}^{-3}$. This simulated stellar mass is also consistent with the constraints ($M_\star\lesssim 10^6\ \rm M_\odot$) from optical and UV observations~\citep{BlackTHUNDER2025}. 

\subsection{Metal Enrichment}\label{subsec:Metal}

\begin{figure}[htb!]
    \centering
    \includegraphics[width=\columnwidth]{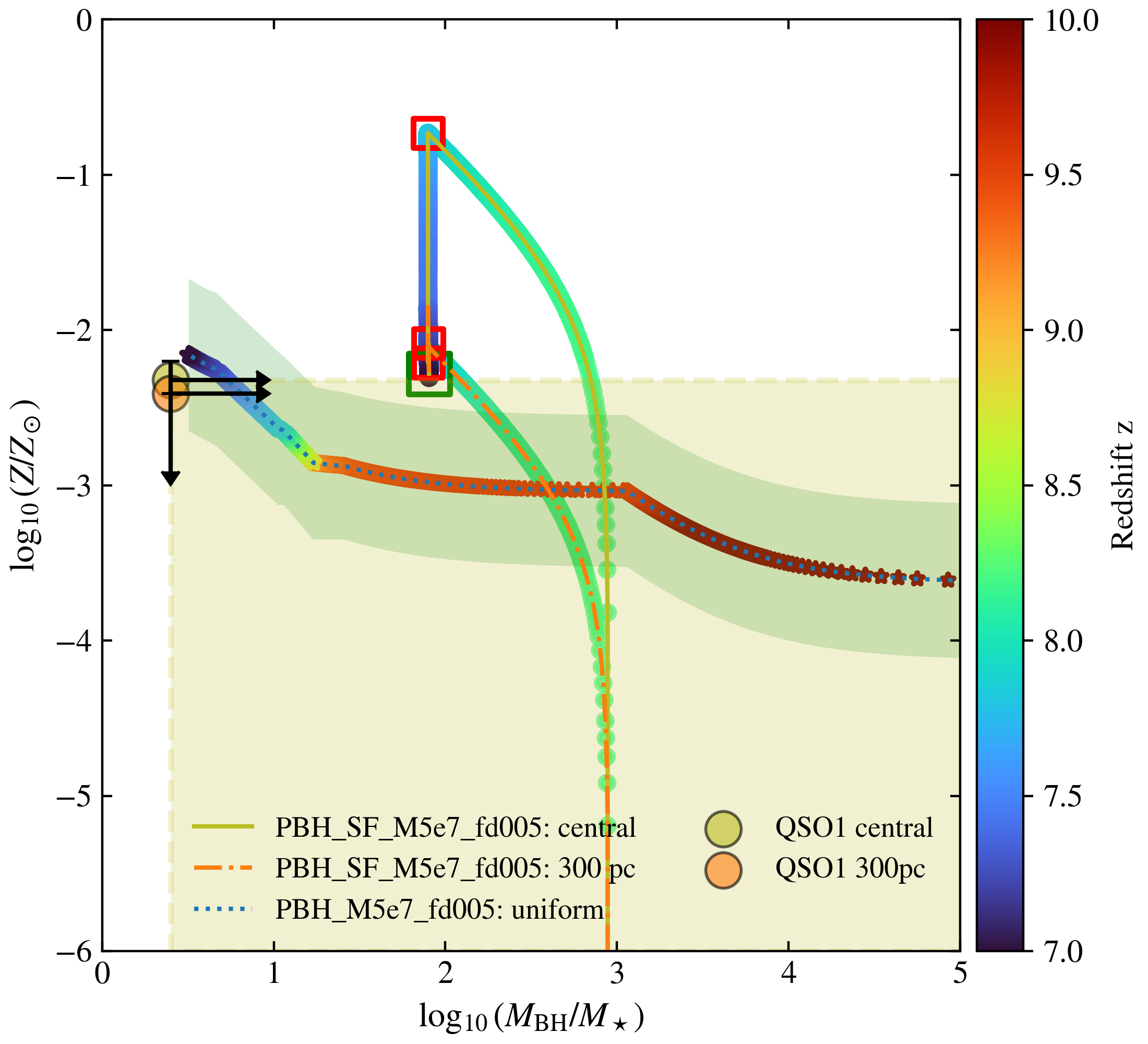}
        \caption{\textbf{Metallicity evolution versus black-hole-to-stellar mass ratio in PBH-seeded galaxies.} Shown are evolutionary tracks from two simulations with different feedback and spatial sampling: \textit{uniform} metallicity measurements (stars with shaded band and dotted blue line) from the \texttt{PBH\_M5e7\_fd005} run, and \textit{central} (circles with olive solid line) and \textit{300~pc} (circles with orange dot-dashed line) apertures from the \texttt{PBH\_SF\_M5e7\_fd005} run (mass-weighted). The color along each curve denotes redshift, from red ($z\sim10$) to blue ($z\sim7$), tracing metal enrichment over time. The green shaded region indicate model dispersion from analytical post-processing of the \texttt{PBH\_M5e7\_fd005} run, while separate tracks show metallicities averaged over $0$–$150$~pc (\textit{central}) and $150$–$300$~pc (\textit{300~pc}) apertures. Square markers highlight the simulation snapshots from \texttt{PBH\_SF\_M5e7\_fd005}: red squares indicate epochs that do not (or only partially) fall within the observationally allowed parameter space for Abell~2744--QSO1, while green squares mark snapshots consistent with the data~\footnote{Note that a few snapshots prior to star formation (not shown) are also found to satisfy the observed data.}. The yellow shaded region marks the allowed parameter space inferred for Abell~2744--QSO1, with its central ($\sim150$~pc) and extended ($\sim300$~pc) metallicity measurements shown as olive and orange points~\citep{Maiolino2025arXiv250522567M}. By $z\approx7$, the simulations naturally reach subsolar metallicities and high $M_{\rm BH}/M_\star$ ratios comparable to those inferred for QSO1.}
    \label{fig:metal_evo}
\end{figure}

In the \texttt{PBH\_SF\_M5e7\_fd005} run, metals are produced by both Pop~III and Pop~II stellar particles primarily during SN feedback events and are distributed, according to our numerical approach, to the gas particles within spherical SN bubbles. 
Once Pop~III stars form within dense gas clouds, their short lifetimes ($\sim3$~Myr) rapidly enrich the local environment, elevating the local metallicity above the critical threshold ($Z/Z_{\odot} \gtrsim 10^{-4}$) and triggering the subsequent formation of Pop~II stars, as shown in the star formation history in Figure~\ref{fig:SFH}. The formation of these dense gas clouds around the PBH simultaneously enhances the accretion rate by a factor of $\sim 10\times$, as evidenced by the correlation between the early SFR burst and the elevated accretion activity in the same figure. During this efficient accretion phase, powerful outflows driven by BH thermal feedback expel metal-enriched gas from the central region, suppressing further star formation while the system continues to accrete pristine, neutral gas from the IGM. This continuous exchange between enriched outflows and pristine inflows drives down the average gas metallicity in the BH's vicinity. The resulting evolution of metallicity as a function of the black-hole-to-stellar mass ratio is shown in Figure~\ref{fig:metal_evo}. 

The \texttt{PBH\_M5e7\_fd005} and \texttt{PBH\_SF\_M5e7\_fd005} runs exhibit similar global enrichment trends but differ in spatial and temporal detail. Prior to the onset of star formation ($z\gtrsim10$), both systems remain essentially pristine, with metallicities below $10^{-6}\,Z_\odot$. Once star formation begins at $z\lesssim10$, Pop~III and subsequently Pop~II stars form rapidly in compact clusters, while supernovae efficiently enrich the surrounding gas. In the \texttt{PBH\_SF\_M5e7\_fd005} run, the central region ($r<150$~pc) enriches most efficiently, reaching $Z/Z_\odot \sim 10^{-1}$ by $z\sim 7.8$, while the outer region ($150$–$300$~pc) attains lower values at $Z/Z_\odot \sim 10^{-2}$. Between $z\sim7$ and 8, strong BH-driven outflows remove enriched gas from the nucleus and suppress further star formation, lowering the average metallicity to levels consistent with Abell~2744--QSO1. The broadened track in Figure~\ref{fig:metal_evo} shows the metallicity evolution inferred from post-processing the \texttt{PBH\_M5e7\_fd005} run, where metallicities are averaged over the full outflow region. The broadening reflects the range of possible metal yields per unit stellar mass from different IMFs, following the methodology of \citet{Maiolino2025arXiv250522567M}. Because this estimate averages over a much larger outflow region, the resulting metallicities are lower due to dilution by metal-poor gas. Overall, both simulations reproduce subsolar metallicities and the extreme $M_{\rm BH}/M_\star$ ratios observed by \textit{JWST}, supporting a picture in which PBH-driven feedback regulates early star formation and chemical enrichment. Abell~2744--QSO1, as observed by JWST, may correspond either to the simulated system during or just prior to its initial starburst at $z\gtrsim9$, or to a brief post-quenching phase near $z\sim7$. Our simulations suggest that strong BH feedback can significantly delay the onset of star formation \citep[see also][]{Zhang2025PBHStar}, 
and that JWST may be capturing an early evolutionary stage of a PBH-seeded system, where the cosmic conditions at $z\sim7$ remain similar to those simulated here at $z\sim9$.

\section{Limitation and Caveats} \label{sec:Caveats}
Before discussing the broader implications of our PBH-seeded scenario, we note several subtleties in comparing our simulations with the observed dynamical constraints of Abell~2744–-QSO1. The inferred point-to-extended mass ratio $M_{\rm point}/M_{\rm extended}\gtrsim2$ reported by \citet{QSO1Direct2025arXiv250821748J} is sensitive to the assumed density profile of the extended component. In our simulations, the steep dark-matter spike formed within $\sim30$~pc of the PBH is observationally indistinguishable from the unresolved BH itself; combining these components increases the effective central mass and yields point-to-extended mass ratios broadly within the allowed range for QSO1. Reproducing the observed rotation curve (\citealt[see their fig.~1]{QSO1Direct2025arXiv250821748J}) within 1$\sigma$, however, requires lowering the central dark matter density (within $\sim 150\ \rm pc$) of the PBH-seeded dark-matter halo by a factor of $\sim2–5$.  

While our current simulations capture the key physical processes linking PBH accretion, stellar feedback, and metal enrichment, several simplifications, although not drastically changing our conclusions, warrant further exploration in future works:
\begin{itemize}
    \item Similar to \citet{Zhang2025PBHStar}, our physical treatment of the PBH seed remains idealized. A single PBH seed is placed in an isolated box, neglecting potentially important effects such as an extended PBH mass function, PBH clustering, or mergers with forming galaxies. A broader PBH mass spectrum could alter small-scale structure formation
    ~\citep[e.g.,][]{Delos2024JCAP...12..005D}, while clustering or early mergers may enhance gas inflow, BH growth, and star formation by supplying additional cold material~\citep[see a review of dual and binary SMBHs in][]{DeRosaSMBHB2019NewAR..8601525D}. Future work will explore a wider range of PBH masses, spatial configurations, simulation box sizes, and accretion prescriptions to quantify their impact on early SMBH–galaxy coevolution.
    \item A related limitation concerns our modeling of the dark matter component. In the present simulations, the PBH grows within collisionless PDM that does not undergo annihilation, self-interactions, or capture by the central PBH. As a result, the halo develops and maintains a steep inner cusp in \texttt{PBH\_SF\_M5e7\_fd005} (as shown in Figure~\ref{fig:profileplot}). This profile is qualitatively consistent with analytic expectations for the DM profile around a central compact object under adiabatic dynamical evolution~\citep[e.g.,][]{Merritt2004PhRvL..92t1304M}. In mixed dark matter scenarios, however, the long-term evolution of the cusp may differ: annihilation can flatten the inner density profile, while PBH–DM capture or dynamical heating can remove dark matter from the center. In our simulations, the formation and evolution of PDM density spikes are related to both the density perturbations developed around the PBH during the radiation-dominated epoch~\citep[see, e.g.,][also captured by the initial conditions of our simulations]{Boucenna2018PBHWIMP, Carr2021MNRAS.506.3648C} and the subsequent in-fall of gas towards the center throughout the cosmic history. The interplay of these processes is still not fully understood. Capturing these processes and comparing to analytical predictions requires a more sophisticated treatment of dark matter microphysics, its interaction with baryons, and BH accretion feedback, which we leave for future work.
    \item Modeling stellar feedback in PBH-seeded systems is intrinsically complex, involving the interplay of radiative heating, SN explosions, and black-hole–driven outflows. In the present implementation, SN feedback is treated in a subgrid manner: each stellar particle releases metals and injects thermal energy into its surroundings at the end of its life. Metals are distributed uniformly in SN bubbles. However, this treatment neglects spatial inhomogeneities and small-scale turbulent mixing during enrichment \citep[e.g.,][]{Ritter_2015}. As a result, it may underestimate local metallicity variations by overestimating the efficiency of chemical homogenization within the ISM~\citep[e.g.,][]{Sarmento2022ApJ...935..174S, Steinwandel2025ApJ...991...16S}. Incorporating explicit metal-diffusion models and (sub-grid) turbulence-driven mixing prescriptions would yield a more physically accurate representation of early chemical enrichment in PBH-seeded galaxies.
    
    \item Our simulations assume BH feedback is purely thermal. However, at later stages, effects such as radiation pressure, mechanical winds, or jet-driven feedback may become important~\citep[see the review by][]{Inayoshi:2020}. These processes can transport metal-enriched gas into the circumgalactic medium more efficiently. 
    A more complete treatment that couples radiative transfer, kinetic feedback, and a multiphase ISM will therefore be essential for capturing the full range of SMBH-driven regulation in early galaxies.
\end{itemize}

In this work, we focus on a proof-of-concept demonstration showing that a massive PBH seed can reproduce the observed properties of Abell~2744–QSO1. If one assumes that the entire JWST “little red dot’’ population originated from a similar PBH-seeded pathway, then using a measured number density of $\sim 4\times 10^{-4}\,\mathrm{cMpc}^{-3}$ at $6.5< z <8.5$ for a rough estimate~\citep[e.g., see table 2 in][]{KokorvLRD2024ApJ...968...38K} would imply an upper limit on the density fraction of $5\times 10^7\ \Msun$ PBH seeds of $f_{\rm PBH}\lesssim 3.2\times 10^{-7}$, much smaller than the existing constraints from dynamical effects and also below typical CMB heating/early-accretion limits~\citep[see, e.g., Sec.~III.E of][]{Carr2021}~\footnote{For $30\,M_\odot<M_{\rm PBH}\lesssim 10^{4}\,M_\odot$, a commonly used approximate scaling for the CMB heating/early-accretion limit is $f_{\rm PBH}\lesssim (M_{\rm PBH}/30\,M_\odot)^{-2}$, with a high-mass flattening at the level of $f_{\rm PBH}\lesssim 10^{-5}$ under Eddington-limited accretion~\citep[see related treatments in][]{Ricotti2008ApJII,Ali-Haimoud2017PhRvD}. On the other hand, CMB $\mu$-distortion constraints tightly limit PBHs at this mass scale~\citep{Nakama2018PhRvD..97d3525N,Hooper2024, DeLuca2025arXiv251219666D}, but these bounds assume Gaussian primordial fluctuations; non-Gaussianity or alternative formation scenarios can substantially weaken them~\citep[see, e.g., section II in][]{EscrivaPBH2024}.}. We note, however, that several recent studies have proposed alternative formation channels for LRDs and QSO1-like systems(e.g.,~\cite{Lu2024PhRvD.109l3016L, Qin2025arXiv250613858Q, Jiang2025SIDM, DeLuca2025arXiv251219666D}, and see section~7 in~\citep{Maiolino2025arXiv250522567M}). Like our study, most of these works remain at the proof-of-concept stage and lack the follow-up numerical modeling required for a quantitative comparison across scenarios. A systematic assessment of competing pathways—spanning population statistics, chemical enrichment histories, environmental/clustering signatures, and multi-wavelength diagnostics—will therefore require substantial future work.

\section{Conclusion}\label{sec:conclusion}

Our simulations reveal a plausible formation pathway for the JWST source Abell~2744--QSO1 within a PBH-seeded scenario. Beginning shortly after the recombination epoch, a massive PBH accelerates the assembly of structure and rapidly accretes the baryons funneled into its seed halo. However, the associated accretion feedback injects substantial thermal energy into the surrounding gas, delaying its cooling and the subsequent onset of star formation. Only once sufficient gas has accumulated and condensed in the potential well does a brief but intense episode of star formation occur in conjunction with enhanced BH accretion. As the accretion rate rises, feedback from the central BH quickly becomes the dominant energy source, heating and expelling gas from the inner regions. This self-regulated phase quenches further star formation and suppresses additional metal enrichment, while powerful outflows redistribute and dilute metals throughout the ISM and circumgalactic medium.

The resulting system—whether observed immediately prior to star formation or shortly after the initial burst—naturally exhibits both low metallicity and an extreme black-hole-to-stellar mass ratio, reproducing the key properties of Abell~2744--QSO1 (as shown in Table~\ref{tab:QSO1_comparison}). Moreover, the predicted PBH accretion rates, primarily regulated by BH feedback in our simulations, are consistent with the value inferred from observations. Early dominance of BH feedback thus provides a compelling physical explanation for the compact, metal-poor, BH-dominated systems uncovered by \textit{JWST}.

\begin{table*}[htb!]
\centering
\caption{Comparison of key observed properties of Abell~2744--QSO1~\citep[adapted from Table 2 in][]{Maiolino2025arXiv250522567M} and the corresponding values from our PBH-seeded simulation (the \texttt{PBH\_SF\_M5e7\_fd005} run) snapshot at $z\simeq 7$.}
\begin{tabular}{lcc}
\hline\hline
Quantity &Observation (QSO1) & Simulation ($z\simeq7$) \\
\hline
BH Mass, $M_{\rm BH}$([$\Msun$])
    & $\sim(2.5$--$10)\times10^{7}$ 
    & $\sim 6\times10^{7}$ \\
Stellar Mass, $M_{\star}$([$\Msun$]) 
    & $\lesssim 2\times10^{7}$ (dynamics), $\sim 10^6$ (optical and UV)
    & $\sim 7.7\times10^{5}$ \\
$M_{\rm BH}/M_{\star}$ 
    & $\gtrsim 2$ (model dependent) 
    & $\sim 80$ \\
Metallicity ($R<150$ pc; [$Z/Z_\odot$]) 
    & $\sim 10^{-2.23} - 10^{-2.44}$ 
    & $\sim 10^{-2.28}$ \\
Metallicity ($150$ pc $<R<300$ pc; [$Z/Z_\odot$]) 
    & $ \lesssim 10^{-2.41}$ 
    & $\sim 10^{-2.31}$ \\
Eddington Ratio, $f_{\rm Edd}$
    & $\sim 0.01$--0.03 
    & $\sim 0.02$ \\
\hline
\end{tabular}
\label{tab:QSO1_comparison}
\vspace{-1ex}
\end{table*}

Distinguishing observationally between a dense, star-free gas cocoon and a compact stellar cluster surrounding a primordial seed will require detailed modeling of the emergent spectral energy distribution (SED), including signatures of massive star formation, nebular line emission, and reprocessed radiation. In practice, however, these phases may be difficult to separate, since the central BH is enshrouded by dense metal-poor gas in both scenarios and the stellar component in our model is subdominant. It is also plausible that \textit{JWST} may capture such systems during the brief, bursty star-formation episodes associated with enhanced BH accretion, as a natural outcome of PBH-regulated growth. Future work will incorporate radiative-transfer tools such as \texttt{CLOUDY}~\citep{CLOUDY2023RMxAA..59..327C} to generate self-consistent SEDs and mock observations, allowing more robust predictions for observational diagnostics. Additional simulations in cosmological environments that permit PBH clustering, interactions, and early mergers will further enable a more complete characterization of the diversity of early SMBH–host galaxy coevolution pathways. In parallel, future \textit{JWST} ultra--deep surveys will be crucial for establishing a statistical census of LRDs. By increasing the number of spatially resolved, high–signal–to–noise detections of low–metallicity, BH–dominated systems, such surveys will enable population–level diagnostics such as number–density estimates, the distribution of $M_{\rm BH}/M_{\star}$, spatial metallicity gradients, and the fraction of sources exhibiting sub–Eddington accretion. These trends will provide key leverage for distinguishing PBH seeds from alternative formation pathways~\citep[see e.g.,][]{Qin2025arXiv250613858Q, Jeon2025arXiv250814155J}, which predict different duty cycles, enrichment histories, and host–environment signatures.

\vspace{-5ex}

\begin{acknowledgments}
 We acknowledge fruitful discussions with Priyamvada Natarajan, Fangzhou Jiang, Kohei Inayoshi, and Barmak Shams Es Haghi and within Mike Boylan-Kolchin's group. The authors acknowledge the Texas Advanced Computing Center (TACC) for providing HPC resources under allocation AST23026. BL gratefully acknowledges the funding of the Royal Society University Research Fellowship and the Deutsche Forschungsgemeinschaft (DFG, German Research Foundation) under Germany's Excellence Strategy EXC 2181/1 - 390900948 (the Heidelberg STRUCTURES Excellence Cluster).
\end{acknowledgments}

%
\vspace{5mm}
\facilities{Lonestar6 and Stampede3 (TACC)}


\software{astropy \citep{2013A&A...558A..33A,2018AJ....156..123A, 2022ApJ...935..167A},  
          Colossus \citep{Diemer2018ApJCOLOSSUS}
          }





\bibliography{Main}{}
\bibliographystyle{aasjournalv7}



\end{document}